# Excess energy and deformation along free edges of graphene nanoribbons

Qiang Lu and Rui Huang*

*Department of Aerospace Engineering and Engineering mechanics, University of Texas, Austin, TX 78712, USA*

*Corresponding author: ruihuang@mail.utexas.edu

Change of the bonding environment at the free edges of graphene monolayer leads to excess edge energy and edge force, depending on the edge morphology (zigzag or armchair). By using a reactive empirical bond-order potential and atomistic simulations, we show that the excess edge energy in free-standing graphene nanoribbons can be partially relaxed by both in-plane and out-of-plane deformation. The excess edge energy and edge force are calculated for graphene nanoribbons with parallel zigzag or armchair edges. Depending on the longitudinal constraint, the compressive edge force leads to either in-plane elongation of the ribbon or out-of-plane buckling deformation. In the former case, the longitudinal strain is inversely proportional to the ribbon width. In the latter case, energy minimization predicts an intrinsic wavelength for edge buckling to be 6.2 nm along the zigzag edge and 8.0 nm along the armchair edge. For graphene nanoribbons of width less than the intrinsic wavelength, interaction between the two free edges becomes significant, leading to anti-phase correlation of the buckling waves.

## I. Introduction

Since isolation of monolayer graphene was first reported in 2005 [1], graphene has drawn tremendous interests for research in physics, materials science, and engineering. To harvest the unique physical properties of monolayer graphene for potential applications in nanoelectronics and electromechanical systems, graphene ribbons with nanoscale widths ($W$ < 20 nm) have been produced either by lithographic patterning [2-4] or chemically derived self assembly processes [5]. The edges of the graphene nanoribbons (GNRs) could be zigzag, armchair, or a mixture of both [6]. It has been theoretically predicted that the special characteristics of the edge states lead to a size effect in the electronic state of graphene and controls whether the GNR is metallic, insulating, or semiconducting [6-8]. Recently, the effects of free edges on the mechanical properties of GNRs have also been studied [9-11], suggesting a size and chirality dependence.

Experimentally it remains a great challenge to observe the atomic structures near free edges of graphene monolayers [12-16]. First-principle calculations based on density functional theory (DFT) have suggested in-plane reconstructions of the edges [17, 18] as a way to relieve the excess edge energy and thus stabilize the planar edge structure. On the other hand, atomistic simulations using empirical potentials have shown rippling, warping, and twisting, with out-of-plane deformation along free edges of graphene monolayers [19-21]. In addition to the excess edge energy, edge stresses and elastic edge moduli have been proposed as continuum thermodynamic properties of the edges [9, 18, 20-22]. However, the reported values for the excess edge energy and the edge stresses are scattered even among different DFT calculations (Table I). In particular, the relative stability of the armchair and zigzag edges has been shown to depend strongly on the choice of the density functional [23]. A couple of previous studies based on empirical potentials predicted edge stresses for the zigzag and armchair edges in opposite orders [20, 21]. To resolve these discrepancies, it is necessary to investigate the physical origins that lead to the excess energy and edge stresses as well as the approximations made in the calculations based on either a particular form of the empirical potential or a particular DFT method.

Table I. Comparison of predicted excess edge energy and edge forces (both in eV/nm) of monolayer graphene.

| | Edge energy ($\gamma$) | | Edge force ($f$) | | $r_0$ |
|---|---|---|---|---|---|
| | Armchair | Zigzag | Armchair | Zigzag | (nm) |
| DFT [17] (GPAW) | 9.8 | 13.2 | - | - | 0.142 |
| DFT [18] (VASP) | 10 | 12 | -14.5 | -5 | 0.142 |
| DFT [22] (SIESTA) | 12.43 | 15.33 | -26.40 | -22.48 | 0.142 |
| MM [20] (AIREBO) | - | - | -10.5 | -20.5 | 0.140 |
| MD [21] | - | - | -20.4 | -16.4 | 0.146 |
| MM (REBO) | 10.91 | 10.41 | -8.53 | -16.22 | 0.142 |

In the present study, using a well-established empirical potential, we show that the change of the bonding environment along the edges leads to excess energy while the mismatch between the edge bonds and the interior bonds induces edge forces. Furthermore, the excess edge energy can be partially relaxed by either in-plane elongation of the graphene nanoribbons or out-of-plane buckling along the edges.

## II. 1-D Relaxation

Based on a reactive empirical bond-order (REBO) potential [24], the chemical binding energy between two carbon atoms takes the form

$$V(r) = V_R(r) - \bar{b} V_A(r), \tag{1}$$

where $r$ is the interatomic distance, $V_R$ and $V_A$ are the repulsive and attractive pair potential functions, respectively, and $\bar{b}$ is the bond-order function that depends on the local bonding environment and multi-body interactions.

Take an infinite, planar graphene monolayer as the reference state. The potential energy per atom is

$$U_0 = \frac{3}{2} V(r_0), \tag{2}$$

where $r_0 = 0.142$ nm is the equilibrium bond length and the corresponding bond-order function $\bar{b} = \bar{b}_0 = 0.9510$. The bond energy at the reference state is: $V_0 = V(r_0) = -4.930$ eV, and thus the energy per atom is: $U_0 = -7.395$ eV.

Now consider an infinitely long ribbon cut from the graphene monolayer with two parallel edges (zigzag or armchair). Assuming no deformation of the graphene lattice for the moment, the change in the bonding environment along the free edges leads to a change in the bond-order function and thus a change of the bond energy. To be specific, the bond-order function of the REBO potential [24] takes the form

$$\bar{b} = b^{\sigma-\pi} + b^{DH} + b^{RC}, \tag{3}$$

where $b^{\sigma-\pi}$ is a function of the bond angles, $b^{DH}$ is a function of the dihedral angles, and $b^{RC}$ represents the influence of radical energetics and $\pi$-bond conjugation. For a planar graphene monolayer at the reference state, $b^{DH} = b^{RC} = 0$ and $b^{\sigma-\pi} = \bar{b}_0$. Along the free edges of a graphene ribbon, however, $b^{RC} \neq 0$ and $b^{\sigma-\pi} \neq \bar{b}_0$, due to the change in the number of neighboring carbon atoms. In particular, for a zigzag edge (Fig. 1a), the bond-order function for the edge bond becomes $\bar{b}_{Z1} = 0.9478$. With the same bond length ($r = r_0$), the bond energy along the zigzag edge is increased from $V_0 = -4.930$ eV to $V_{Z1} = -4.858$ eV. Consequently, the energy per atom in the first row of the zigzag edge increases from $U_0 = -7.395$ eV to $U_{Z1} = -4.858$ eV. Note that each edge atom is associated with two edge bonds instead of three. In addition, the energy per atom in the second row of the zigzag edge also increases because of association with the edge bonds: $U_{Z2} = (2V_{Z1} + V_0)/2 = -7.323$ eV. Together, relative to the reference state, the excess energy per unit length of the zigzag edge is

$$\gamma_Z = \frac{1}{\sqrt{3} r_0} \left( U_{Z1} + U_{Z2} - 2U_0 \right), \tag{4}$$

which gives 10.61 eV/nm or 1.70 nN by the REBO potential.

Similarly, for an armchair edge (Fig. 1b), the bond energy changes in the first and second rows: $V_{A1} = -5.268$ eV and $V_{A2} = -4.858$ eV. The energy per atom is then: $U_{A1} = (V_{A1} + V_{A2})/2 = -5.063$ eV in the first row and $U_{A2} = (V_{A2} + 2V_0)/2 = -7.359$ eV in the second row, both greater than the reference value. The excess energy per unit length of the armchair edge is thus

$$\gamma_A = \frac{2}{3 r_0} \left( U_{A1} + U_{A2} - 2U_0 \right), \tag{5}$$

which is 11.12 eV/nm or 1.78 nN, slightly higher than that of the zigzag edge.

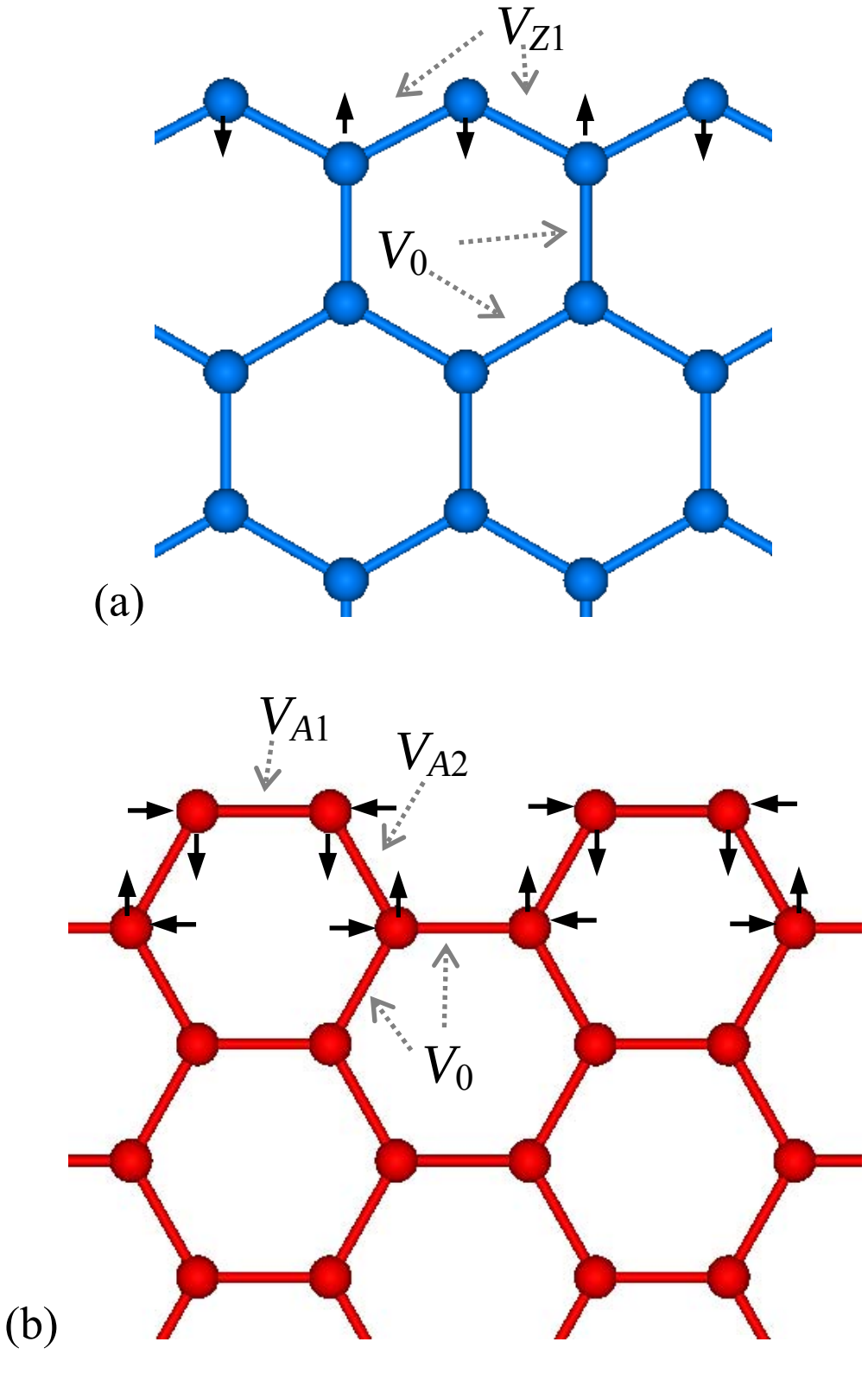

Figure 1. Illustration of the bond structures, bond energies, and unbalanced interatomic forces at the un-relaxed zigzag (a) and armchair (b) edges. Arrows indicate directions of unbalanced forces acting on the edge atoms.

The above analysis of the excess edge energy has assumed no deformation or bond reconstruction along the edges, termed as *un-relaxed* edges. However, the change of the bonding environment also leads to changes in the equilibrium bond length and bond angles along the edges.

As a result, the interatomic forces acting on each atom are not balanced along the un-relaxed edges. As illustrated in Fig. 1, the forces acting on the atoms in the first two rows of the un-relaxed zigzag edge are un-balanced in the direction perpendicular to the edge, while the unbalanced forces are in both perpendicular and parallel directions for atoms along the un-relaxed armchair edge. On the other hand, the interatomic forces acting on each atom are balanced in the parallel direction along the zigzag edge due to symmetry. Consequently, the edge atoms tend to displace in the direction of the unbalanced forces, leading to spontaneous deformation of the graphene lattice near the free edges and relaxation of the excess edge energy. Using the REBO potential, we simulate the edge relaxation by a standard Molecular Mechanics (MM) method, minimizing the total potential energy in a graphene ribbon with two parallel free edges.

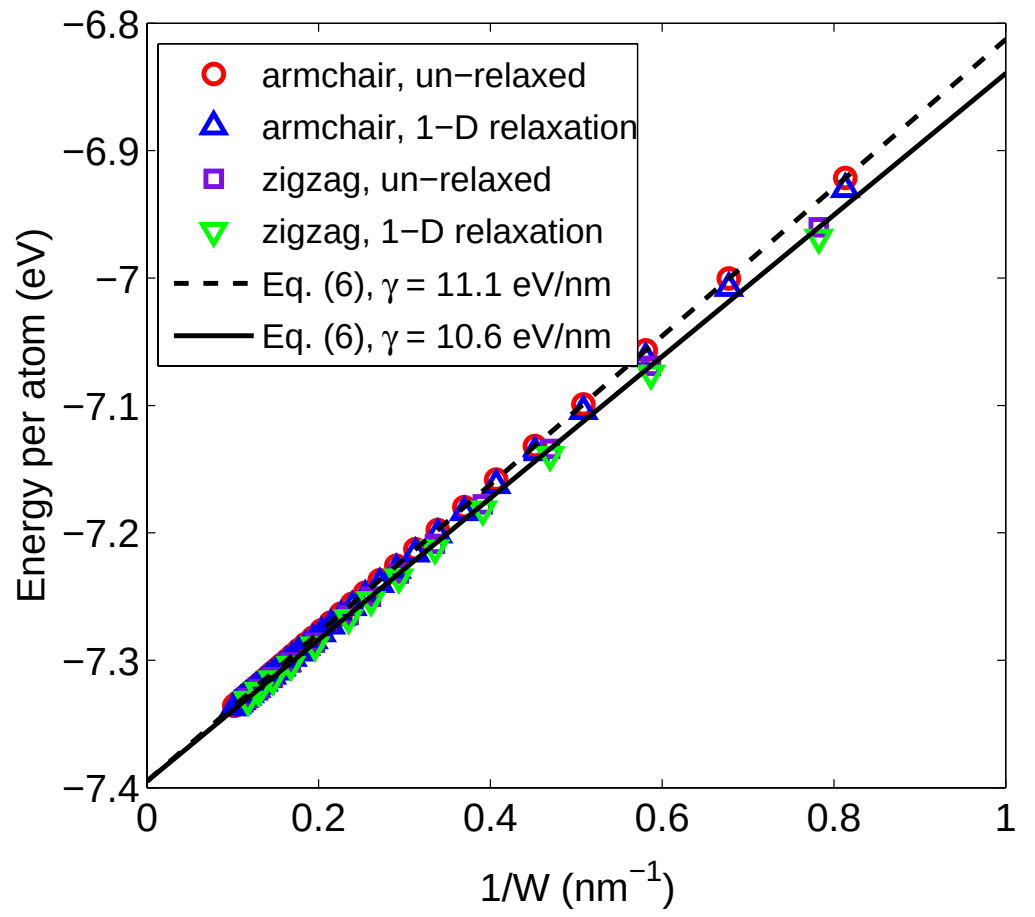

Figure 2. Average energy per atom of graphene nanoribbons as a function of the ribbon width, with un-relaxed edges and after 1-D edge relaxation.

For comparison, we first calculate the average potential energy per atom with un-relaxed edges, for which the lattice structure of each ribbon is taken directly from the reference state of a fully relaxed, planar graphene monolayer. As shown in Fig. 2, the potential energy per atom increases linearly with the inverse of the ribbon width, which can be understood as a result of the excess edge energy, namely,

$$\bar{U}(W) = U_0 + \frac{2\gamma}{N} = U_0 + \frac{S_0}{W}\gamma, \qquad (6)$$

where $\bar{U}$ is the average energy per atom, $W$ is the ribbon width, $\gamma$ is the excess edge energy per length, $S_0 = \frac{3}{2}\sqrt{3}r_0^2$ is the area of the unit cell of graphene (containing two carbon atoms), and $N = 2W/S_0$ is the number of carbon atoms per unit length of the ribbon. Equation (6) reveals the dependence of the average energy on the ribbon width, which agrees closely with the atomistic calculations for the un-relaxed edges (Fig. 2).

Next, in the MM simulations, we allow the atoms to move in the direction that reduces the total potential energy by a quasi-Newton algorithm. Only in-plane displacements of the atoms are allowed for the moment. Periodic boundary conditions are assumed at both ends of the graphene ribbons, with the end-to-end distance fixed. Upon such relaxation, the ribbon width shrinks slightly, while the ribbon length does not change, thus termed as *1-D relaxation*. Figure 2 shows that the average energy per atom in each ribbon decreases slightly after the 1-D relaxation. By Eq. (6), the excess edge energy after the 1-D relaxation is calculated from the average energy and plotted in Fig. 3. Before relaxation, the excess edge energies agree closely with those predicted by Eqs. (4) and (5), for the zigzag and armchair edges, respectively. After 1-D relaxation, both the edge energies are reduced by roughly 2%, i.e., $\gamma_Z = 10.41$ eV/nm and $\gamma_A = 10.91$ eV/nm. The excess edge energy is independent of the ribbon width before and after 1-D relaxation.

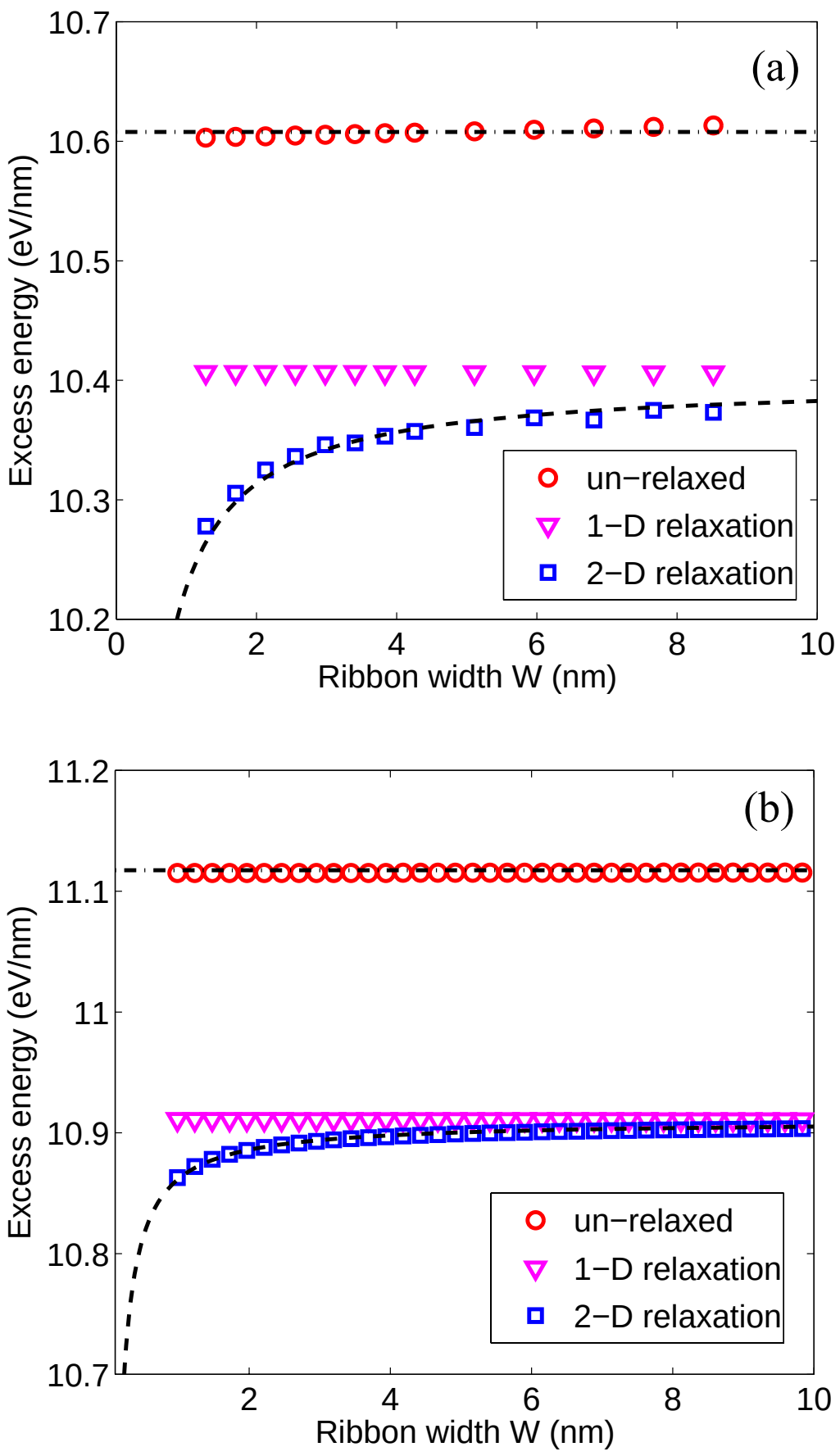

Figure 3. Excess energy per unit length vs ribbon width: (a) zigzag edge; (b) armchair edge. The dashed lines are analytical predictions (Eq. 7) for 2-D relaxation based on the edge energy and edge force from 1-D relaxation.

Upon 1-D relaxation, the interatomic forces acting on each atom are balanced in all directions. However, the mismatch between the equilibrium bond length at the edges and that at the interior of the graphene ribbon leads to a compressive internal force along the free edges, which was called *edge stress* previously [18, 20-22]. The internal edge forces are self-balanced in an infinitely long ribbon, as illustrated in Fig. 4. To evaluate the magnitude of the edge force, we calculate the total internal force acting on a cross section of the graphene ribbon (with 2 parallel edges) after the 1-D relaxation, which equals twice the corresponding edge force. This calculation gives the edge forces: $f_Z$ = -16.22 eV/nm or -2.60 nN for the zigzag edge and $f_A$ = -8.53 eV/nm or -1.36 nN for the armchair edge, both compressive as indicated by the negative sign. Alternatively, the edge forces can be determined from variation of the excess edge energies in strained graphene ribbons [20, 22]. As will be discussed in Section III, both methods predict essentially the same edge forces.

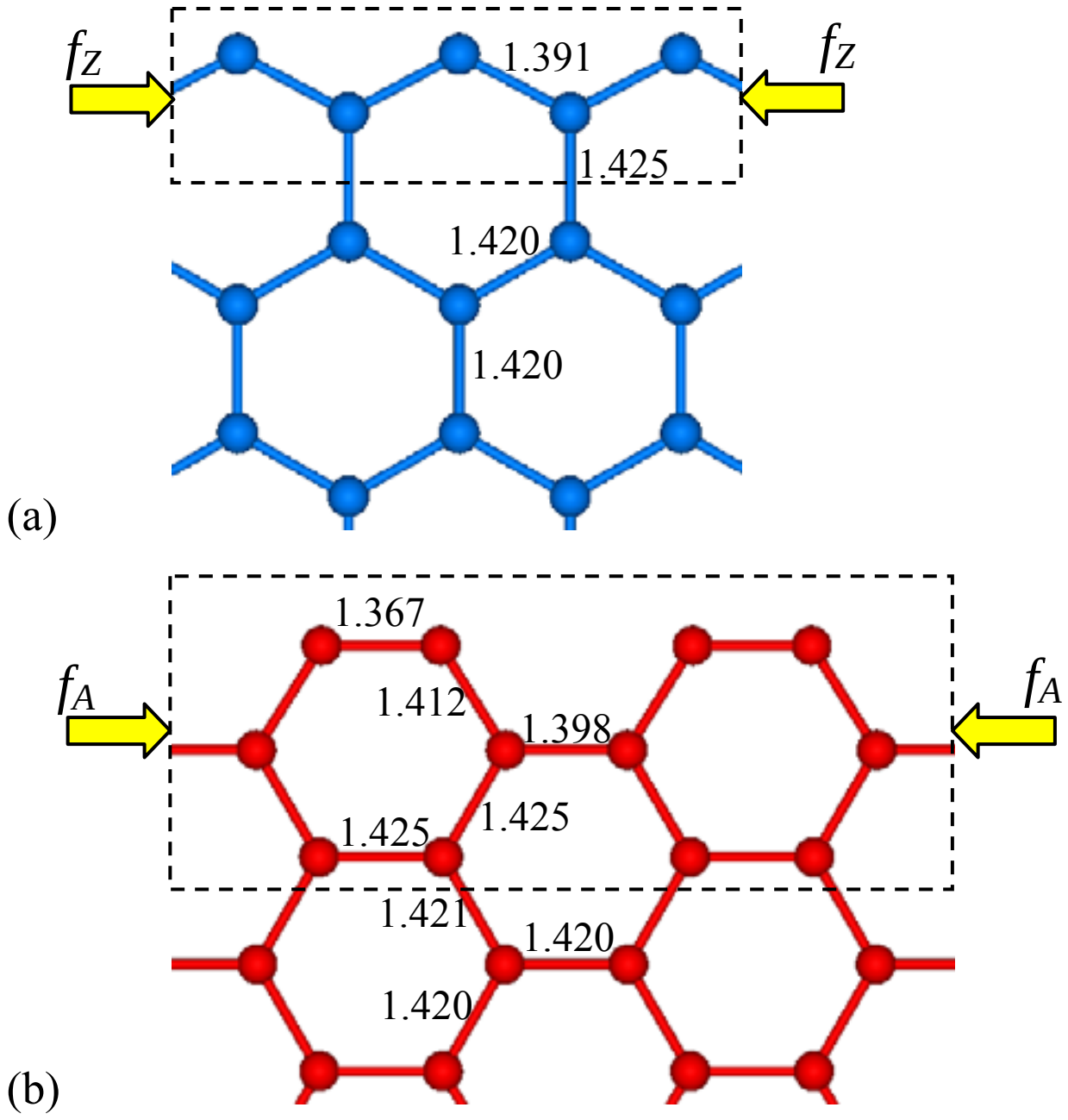

Figure 4. The bond structures (with bond lengths in Å) at zigzag (a) and armchair (b) edges after 1-D relaxation. The dashed rectangular boxes represent effective edge layers that are subjected to compressive edge forces.

It is noted that, while the excess edge energies for the zigzag and armchair edges differ slightly (< 5%), the edge force of the zigzag edge is nearly twice that of the armchair edge. Table I compares the edge energies and edge forces from the present study (after 1-D relaxation) with other calculations. Several density functional theory (DFT) calculations have predicted similar excess edge energies [17, 18, 22]. Notably, the excess energy for the zigzag edge (without reconstruction) from the DFT calculations is higher than that for the armchair edge, opposite to the predictions by the REBO potential. Experimentally both armchair and zigzag edges have been observed in graphene monolayers [13-16], suggesting small energy difference between the two edges. Recently, in situ TEM observations have reported mostly zigzag edges at the sublimation fronts [15, 16] and conversion of armchair edges to zigzag edges [14], suggesting a lower excess energy and thus better stability of the zigzag edge. Although reconstruction of the zigzag edge was predicted by the DFT calculations to have a lower excess energy than that of the armchair edge [17, 18], the type of edge reconstruction has not been confirmed experimentally. In the present MM simulations, no edge reconstruction is observed.

For the edge forces, the DFT calculations have predicted quite different values among themselves [18, 22], possibly due to the uses of different approximations and methods. On the other hand, edge forces similar to the present MM calculations were predicted previously using a different empirical potential [20], while a molecular dynamics (MD) simulation using an earlier version of the REBO potential predicted considerably larger edge forces [21]. The differences may result from different equilibrium bond lengths of graphene predicted by the different empirical potentials, as listed in Table I. The REBO potential used in the present study predicts an equilibrium bond length in close agreement with the DFT calculations. Despite the discrepancies, all calculations have predicted compressive edge forces for both the zigzag and armchair edges, which lead to further deformation and relaxation along the edges, as discussed in the subsequent sections.

## III. 2-D Relaxation

Due to the compressive edge forces, the total potential energy in a graphene ribbon can be partially relaxed either by elongation in the longitudinal direction of the ribbon (namely, 2-D relaxation) or by out-of-plane displacement of the atoms (edge buckling, Section IV). To simulate the 2-D relaxation, while only in-plane displacements of the atoms are allowed, the end-to-end distance of the graphene ribbon is varied gradually to impose a longitudinal strain ($\varepsilon$) until the total potential energy reaches a minimum. Figure 5 shows representative results from the MM simulation for 2-D relaxation of a graphene ribbon with zigzag edges.

The average energy per atom in the graphene ribbon is calculated, which reaches a minimum at a positive longitudinal strain, $\varepsilon \sim 0.005$. In addition, the total force ($F$) acting at each end of the graphene ribbon is evaluated, which is compressive at zero strain and becomes zero at the same strain for the minimum energy. The results can be understood as follows. First, the energy per atom of the graphene ribbon can be written as a function of the imposed longitudinal strain:

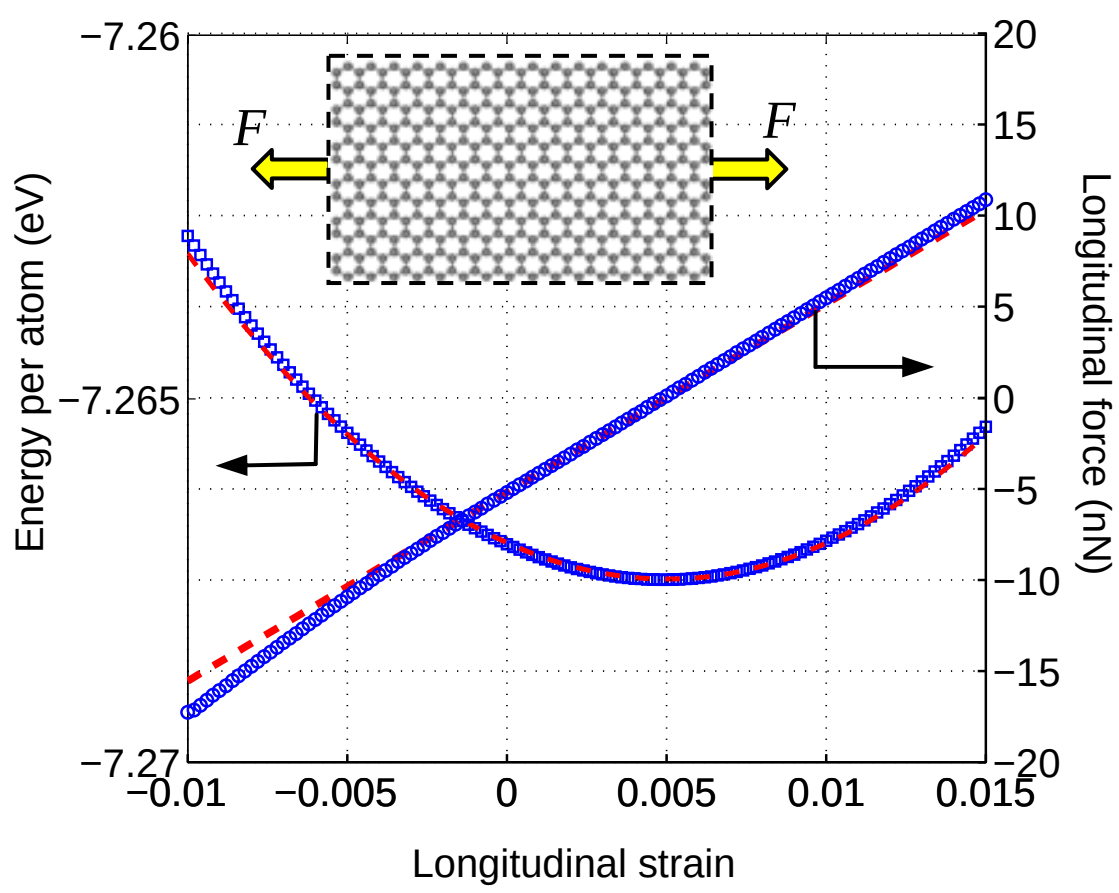

Figure 5. Average energy per atom and longitudinal force versus imposed longitudinal strain for a graphene ribbon with zigzag edges. The ribbon width $W$ = 3.8 nm. The dashed lines are predicted by Eqs. (7) and (8) for the energy and force, respectively.

$$\bar{U}(\varepsilon, W) = U_0 + \frac{S_0}{W}\gamma(\varepsilon) + \frac{S_0}{4}Y\varepsilon^2, \tag{7}$$

where $Y$ is the 2D Young's modulus of the monolayer graphene ($Y$ = 243 N/m by the REBO potential [25]), and $\gamma = \gamma(\varepsilon)$ is the excess edge energy taken as a function of the strain. The third term on the right-hand side of Eq. (7) accounts for the interior elastic strain energy of the graphene ribbon in addition to the edge energy. Here we assume linear elastic behavior of graphene under infinitesimal strain ($\varepsilon << 1$).

As illustrated by the inset of Fig. 5, the work done by the longitudinal force equals the increase of the total potential energy, namely, $N\delta\bar{U} = F\delta\varepsilon$, where $N = 2W/S_0$ is the number of carbon atoms per unit length of the ribbon. Thus, the longitudinal force is

$$F = N\frac{d\bar{U}}{d\varepsilon} = YW\varepsilon + 2\frac{d\gamma}{d\varepsilon}. \tag{8}$$

In the special case of 1-D relaxation, we have $\varepsilon = 0$ and thus $F = 2(d\gamma/d\varepsilon)_{\varepsilon=0}$. On the other hand, we have $F = 2f$ by force balance between the internal edge force ($f$) and the external force. Therefore, the edge force after the 1-D relaxation is related to the edge energy function as $f = (d\gamma/d\varepsilon)_{\varepsilon=0}$. Assuming infinitesimal strain, the edge energy function is approximately

$$\gamma(\varepsilon) \approx \gamma_0 + f\varepsilon, \tag{9}$$

where $\gamma_0$ is the excess edge energy after 1-D relaxation as those obtained in the previous section for the zigzag and armchair edges, respectively. Previous works included a quadratic term in the expansion [9, 20], which is found to have negligible effect (except for very narrow ribbons) and thus ignored in the present study.

By inserting Eq. (9) into Eq. (7), we see clearly that, while the interior elastic strain energy increases with the longitudinal strain, the excess energy along the edges decreases linearly with positive strain due to the compressive energy force ($f$ < 0). The competition between the two strain-dependent terms leads to a minimum energy at a positive strain (elongation):

$$\varepsilon = -\frac{2f}{YW}, \tag{10}$$

at which point the corresponding force vanishes ($F$ = 0). As shown by the dashed lines in Fig. 5, by using the values after 1-D relaxation for $\gamma_0$ and $f$ in Eq. (9), the energy and the force predicted by Eqs. (7) and (8) agree closely with the MM results at small strains. This confirms that the edge forces evaluated by force calculations in the previous section are identical to those based on Eq. (9); the latter was used in several previous studies [20, 22].

Figure 6 plots the longitudinal strain corresponding to the minimum energy in graphene ribbons with different ribbon widths ($W$). As predicted by Eq. (10), the longitudinal strain after 2-D relaxation is inversely proportional to the ribbon width. By substituting Eq. (10) into Eq. (7), we obtain an apparently width-dependent excess energy after 2-D relaxation, as shown in Fig. 3.

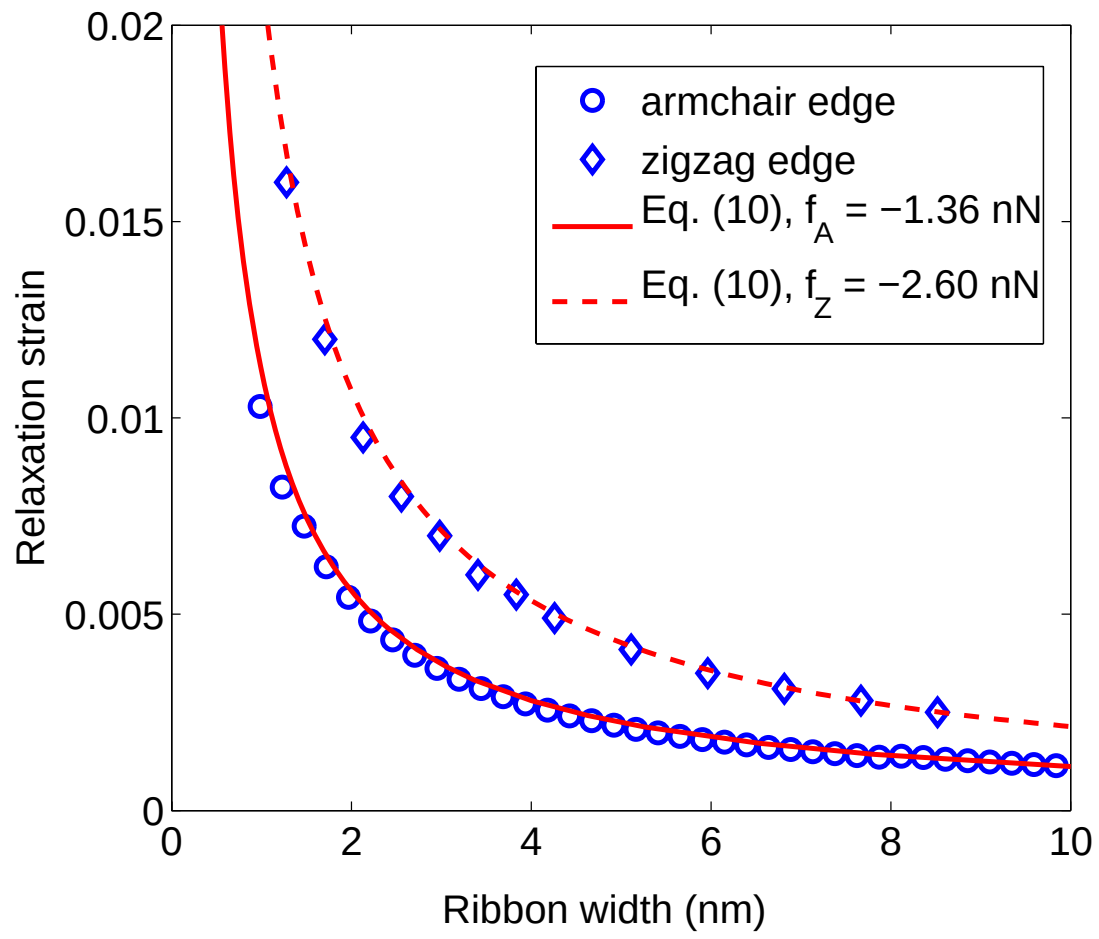

Figure 6: Longitudinal strain of graphene nanoribbons after 2-D relaxation: comparing MM calculations (open symbols) and the predictions by Eq. (10).

## IV. Edge Buckling

An alternative mode of edge relaxation can be achieved by three-dimensional deformation of the graphene ribbon. The compressive edge force motivates out-of-plane buckling along the free edges, but opposed by the bending stiffness of graphene [26]. The competition leads to an intrinsic wavelength for the edge buckling. Figure 7 shows two examples of graphene ribbons with edge buckling by MM simulations. The color indicates the magnitude of out-of-plane displacements of the atoms. Clearly, the buckle amplitude maximizes along the free edges and decays away from the edges. Similar edge buckling was predicted previously [18-21].

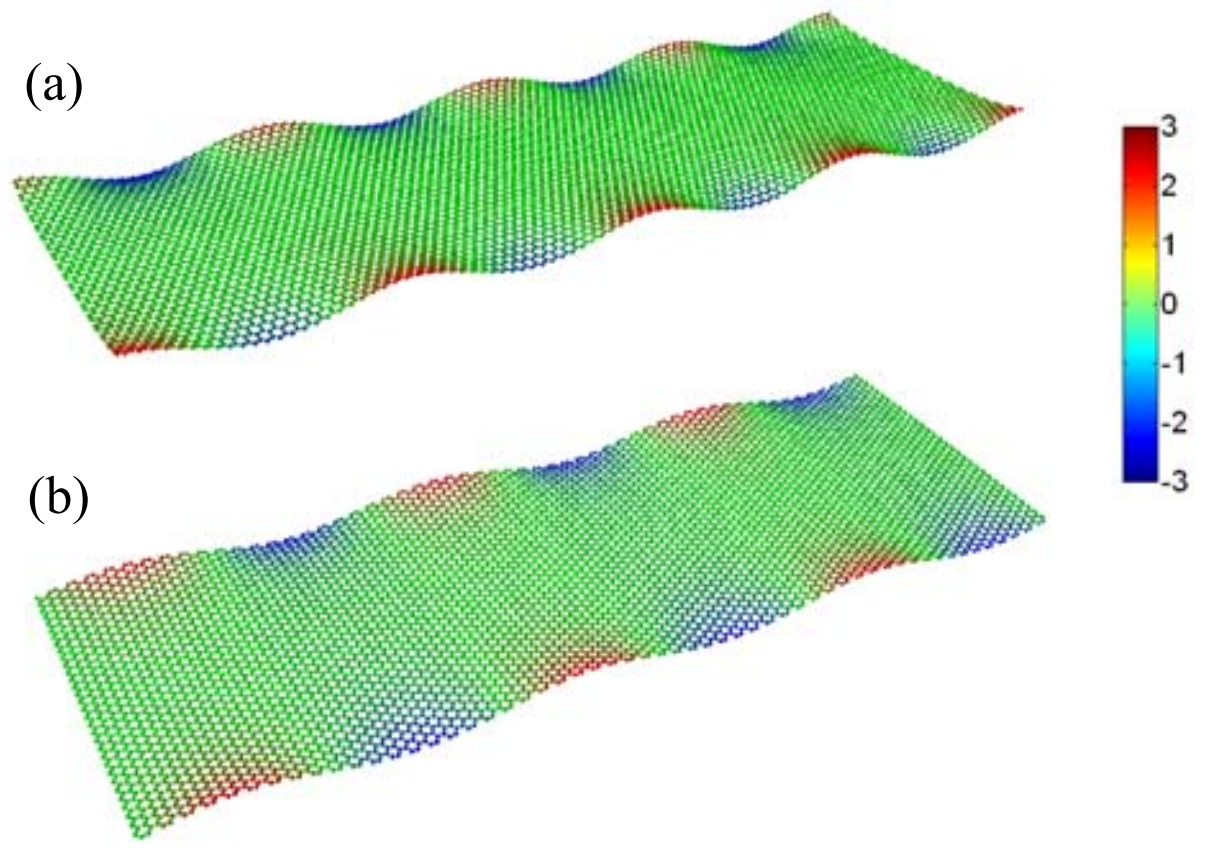

Figure 7. Edge buckling of graphene nanoribbons with (a) zigzag edges ($W$ =7.7 nm, $L$ = 23.6 nm) and (b) armchair edges ($W$ = 7.9 nm, $L$ = 23.4 nm). Color bar indicates the out-of-plane displacement (in Å).

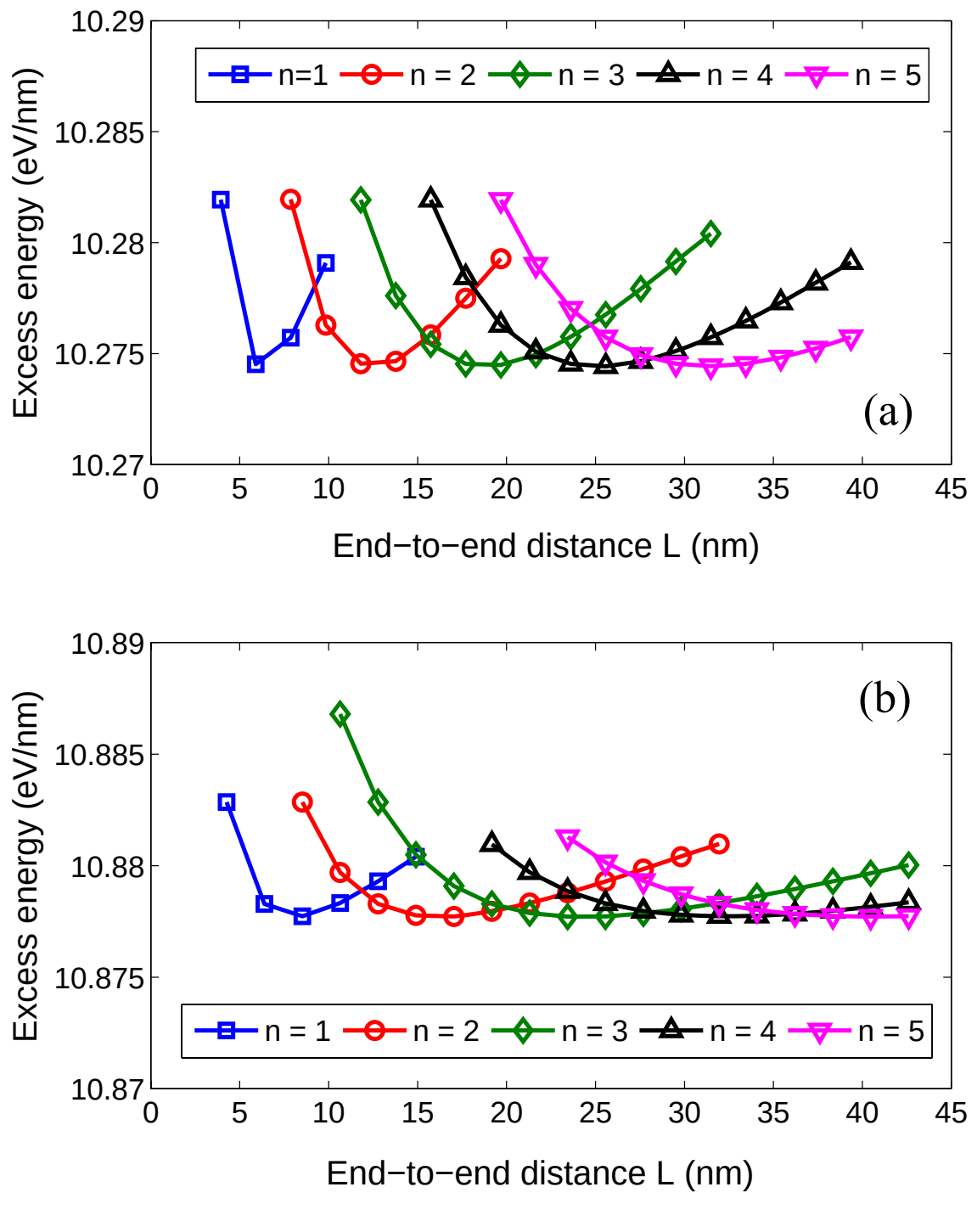

Figure 8. Excess energy of graphene nanoribbons with edge buckling for different wave numbers. (a) zigzag edge ($W$ = 7.7 nm); (b) armchair edge ($W$ = 7.9 nm).

Figure 8 shows the excess energy calculated by MM simulations of graphene ribbons with edge buckling. For each graphene ribbon, the end-to-end distance is fixed during the MM simulation along with the periodic boundary conditions at both ends, and out-of-plane perturbations are introduced to trigger the buckling deformation. The excess energy is calculated as the total energy increase per unit length of the free edges relative to the ground state of graphene, thus including both the edge energy and the interior bending energy of the ribbon. It is found that the excess energy depends on both the end-to-end distance ($L$) and the buckling wave number ($n$). For each $L$, different buckling modes are obtained from the MM simulations, indicating that more than one local energy minimum exists. Figure 9 re-plots the excess energy as a function of the buckle wavelength, $\lambda = L/n$, in which the MM results in Fig. 8 collapse onto one single curve and the excess energy minimizes at a particular wavelength for each edge configuration. Therefore, an intrinsic wavelength for edge buckling exists, which is 6.2 nm for the zigzag edge and is 8.0 nm for the armchair edge, according to Fig. 9.

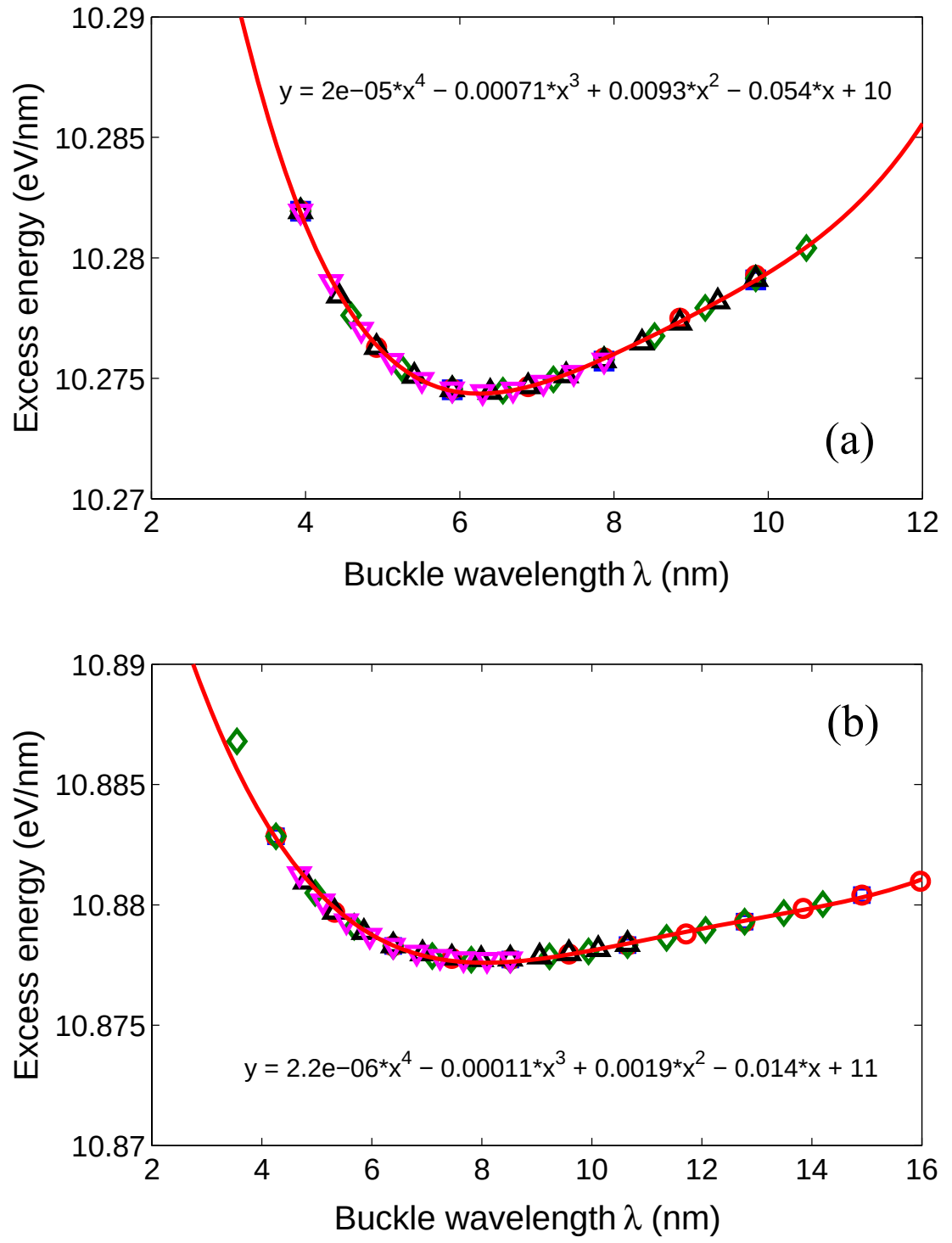

Figure 9. Excess energy of graphene nanoribbons with edge buckling. (a) zigzag edge ($W$ = 7.7nm); (b) armchair edge ($W$ = 7.9 nm). The MM results are fitted by 4$^{th}$-order polynomial functions.

Previous studies [19-21] have reported wavelengths of edge buckling in a wide range (3-10 nm). A continuum model based on elastic plate theory was used to determine the scaling laws for the buckle amplitude and penetration depth as the buckle wavelength varies, where bending stiffness of the graphene was ignored and no intrinsic wavelength was determined for either zigzag or armchair edge [20]. On the other hand, a standard linear perturbation analysis reportedly predicted the buckle wavelength to scale with the ratio between the bending stiffness and the edge force, i.e., $\lambda \approx -4\pi^2 D / f$ [21]. Using the edge forces obtained from the present study along with the bending stiffness $D$ = 0.225 nN-nm [26], the buckle wavelength would be 3.4 nm for the zigzag edge and 6.5 nm for the armchair edge, both considerably shorter than the MM results. It is thus necessary to develop a more rigorous continuum model for edge buckling, which should take into account both the bending stiffness and in-plane stiffness of graphene as well as the effect of bending curvature on the edge energy.

We have performed MM simulations for graphene nanoribbons with different widths and found that the wavelength of edge buckling is independent of the ribbon width for wide ribbons with $W$ > 6 nm. However, for narrower ribbons, the buckle wavelength changes slightly with the ribbon width, most likely due to the proximity of the two edges that interact with each other. It is also noted that the edge-edge interaction in narrow graphene ribbons ($W$ < 6 nm) can lead to anti-phase correlation of the buckling waves along the two parallel edges (Fig. 10), in which case the nanoribbon appears to be twisted like those reported in previous studies [20, 21].

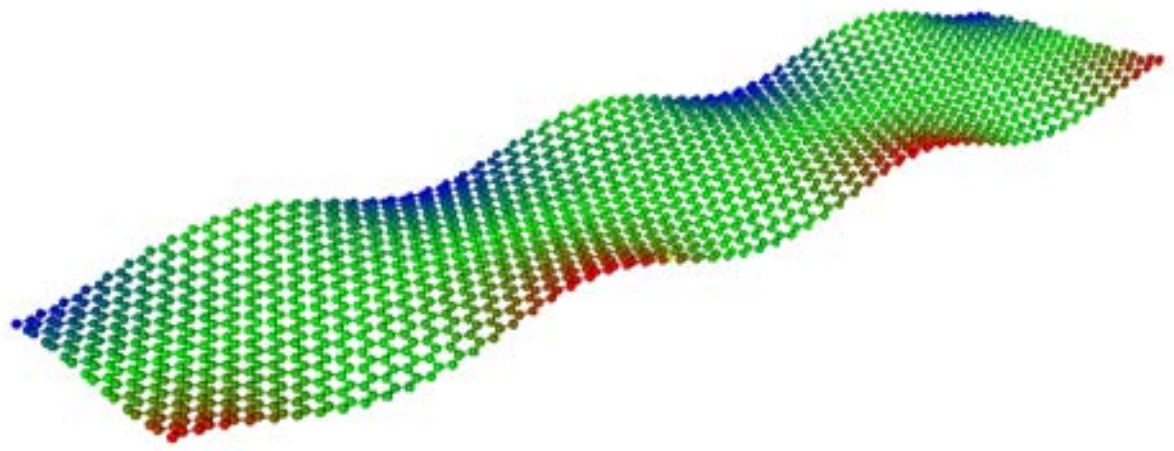

Figure 10. Anti-phase correlation of edge-buckling in a narrow graphene ribbon ($W$ = 3.4 nm and $L$ = 14.8 nm).

## V. Summary

We have analyzed excess energy and induced deformation along zigzag and armchair edges of graphene nanoribbons by atomistic simulations using a reactive empirical bond-order (REBO) potential. The in-plane 1-D relaxation of the edges leads to an edge energy (10.91 eV/nm for armchair and 10.41 eV/nm for zigzag) and a compressive edge force (–8.53 eV/nm for armchair and –16.22 eV/nm for zigzag). The compressive edge force motivates elongation of the graphene ribbon in the 2-D relaxation and out-of-plane buckling along the edges. An intrinsic wavelength for edge buckling is predicted to be 8.0 nm for the armchair edge and 6.2 nm for the zigzag edge. We note that the atomistic simulations in the present study are limited to equilibrium states at temperature $T$ = 0 K, and effects of finite temperature are expected for both the excess edge energy and the edge forces as well as the induced deformation of graphene nanoribbons.

## Acknowledgments

The authors gratefully acknowledge funding this work by National Science Foundation through Grant No. CMMI-0926851.